\documentstyle[12pt]{article}
\begin{document}
\begin{flushright}
OCHA-PP-107\\
NDA-FP-43\\
May (1998)
\end{flushright}

\begin{center}

{\large\bf\sc A Rule of Thumb Derivation
of Born-Infeld Action\\ for D-branes}\\
\vspace{1cm}
RIKA ENDO, RIE KURIKI, SHIN'ICHI NOJIRI$^\sharp$, and AKIO SUGAMOTO\\
\medskip {\it Department of Physics, Ochanomizu University \\ 2-1-1,
Otsuka, Bunkyo-ku, Tokyo 112, Japan \\ \smallskip $\sharp$ Department of
Mathematics and Physics, National Defence Academy\\ Hashirimizu, Yokosuka,
239, Japan}
\end{center}
\bigskip
\begin{abstract}

\bigskip       

A rule of thumb derivation of the Dirac-Born-Infeld action for D-branes is
studied $\grave{a}~ la$ Fradkin and Tseytlin, by simply integrating out of
the superstring coordinates in a narrow strip attached to the D-branes.  In
case of superstrings, the coupling of Ramond-Ramond fields as well as the
Dirac-Born-Infeld type coupling of the Neveu Schwarz-Neveu Schwarz fields
come out in this way.   

\end{abstract}
\section{Introduction}

 Recently, solitonic extended 
objects have been found in string theories and they
are called D-branes~\cite{Polchinski}. (``D" comes from the Dirichlet boundary
condition.)
Investigation of the dynamics of D-branes is very important in order to
learn the nonperturbative behavior of string theories in the strong
coupling region. It is
also known that the interaction of D-brane with massless (bosonic) modes of
superstrings is described by a kind of the Born-Infeld
action~\cite{F-T}-\cite{Scmid}.

Let us recapitulate the various derivations of the Born-Infeld action in
string theories.
In Ref.\cite{F-T} the authors have studied the effective action for an Abelian
vector field coupled with the open bosonic string in the space-time
dimensions $D=26$, and derived the Born-Infeld type action in a constant
background field strength.  
For open superstrings with Dirichlet (fixed) boundary conditions, boundary
states have been introduced, and the infinities and anomalies as well as
their relationship to the moduli space boundaries have been analyzed in
detail\cite{Cai}.  
Through the investigation of the boundary states of open superstrings, the
authors in Ref.\cite{Callan} have found the Born-Infeld type effective action,
from which loop corrected field equations in the $\sigma$-model approach
have resulted.

On the other hand, in Ref.\cite{Leigh}, a new class of $\sigma$-models is
introduced, which is characterized by the choice of boundary conditions for
strings. 
In particular, Leigh has shown that the effective action, which is  consistent
with the vanishing condition of open string's $\beta$ functions, is the
Dirac-Born-Infeld action. 
In Ref.\cite{Douglas} a set of universal couplings between Ramond-Ramond (R-R)
gauge fields of superstrings and the internal gauge field of D-branes have
been determined as a geometrical consequence.
In Ref.\cite{Li} the massless sector of boundary states of D-branes is
constructed, and the picture changing between various vertex operators is
studied.
After sensibly summarizing the Born-Infeld actions for D-branes, the
self-duality for type IIB superstring as well as the duality relationship
between type IIA superstring and  the 11-D supermembrane are investigated 
using the Born-Infeld actions \cite{Scmid}.

Here we will study a rule of thumb derivation of the Born-Infeld action. 
It is a way to understand the derivations of the Born-Infeld action $\grave{a}~ la$ 
Fradkin and Tseytlin, by simply integrating
out of the superstring coordinates in a narrow strip attached to the
D-branes.  The way of thinking is easy to follow and as a result it may be useful for many reserchers.

\section{Born-Infeld Action for Bosonic String} 
We first start with the bosonic string case, where the open string
described by the coordinate $X^{\mu}(\tau, \sigma)$ is coupled with the
dilaton $\phi(X)$, the graviton $G^{\mu\nu}(X)$, and the anti-symmetric
tensor field (the two-form field of the Neveu Schwarz-Neveu Schwarz (NS-NS)
sector) $B^{\mu\nu}(X)$. The open string has an end point which is stuck to
the D-brane, but can move freely inside the $p$ dimensionally extended
D-brane (D-p brane). The position of the D-brane may be given by the
coordinates $(x^0, \cdots, x^p, x^{p+1}, \cdots, x^{9})$. Here,
$x^{m}~(m=0, \cdots ,p)$ are the inner coordinates, whereas
$x^{i}=Y^{i}(x^0, \cdots, x^p) (i=p+1, \cdots ,9)$ are the outer
coordinates of the D-brane.

Then, we have the following action,
\begin{eqnarray}
S &=& \frac{1}{4\pi}\int d^2\xi \sqrt{g(\xi)} R^{(2)} \phi(X) \nonumber \\
&+& \frac{1}{\kappa} \int d^2 \xi \left( \eta^{ab} G_{\mu\nu} (X) +
\epsilon^{ab} B_{\mu\nu}(X) \right)\partial_{a} X^{\mu} \partial_{b}
X^{\nu} \nonumber \\
&-& \frac{1}{\kappa} \int d\tau \left( A_m(X) \partial_{\tau} X^m (\tau,
\sigma=0) - N_i(X) \partial_{\sigma}X^i(\tau, \sigma=0)
\right).\label{bosonic action}
\end{eqnarray}
Here the end point trajectory of the open string is specified simply by
$\sigma=0$, but the other end  point $\sigma=\pi$ is also possible to stick
to the same D-brane.  The trajectory may couple with the two kinds of
vector fields, $A_{\mu}(X)$ and $N_{\mu}(X)$, and $\kappa^{-1}$ represents
the energy stored inside the unit length of the string, being related to
the Regge slope $\alpha'$ as $\kappa=2\pi~\alpha'$.

Since the worldsheet of the open string (in the relevant strip region) is
a disc, the Euler number appearing in front of the dilaton field is unity,
$\chi=\frac{1}{4\pi}\int d^2\xi \sqrt{g(\xi)} R^{(2)}(\xi)=1$. In the low-energy 
effective theory, all the external fields vary slowly compared with
the rapid oscillation of the string modes, so that the $X^{\mu}$ can be
replaced by its center-of-mass value $x^{\mu}$. 

By using the Stoke's and Gauss' theorems, the action $S$ becomes 
\begin{equation}
\label{action}
S= \phi(x) + \frac{1}{\kappa} \int d^2 \xi~M_{\mu\nu} (x)~ ( \eta^{ab} +
\epsilon^{ab} ) \partial_{a} X^{\mu} \partial_{b} X^{\nu}, \end{equation}
where
\begin{eqnarray}
M_{\mu\nu} (x)&=&G_{\mu\nu} (x) + N_{\mu\nu} (x) + B_{\mu\nu} (x) +
F_{\mu\nu} (x), \\
N_{\mu\nu} (x)&=&\partial_{\mu} N_{\nu} +\partial_{\nu} N_{\mu},~~
\mbox{and}~~ F_{\mu\nu} (x)=\partial_{\mu} A_{\nu} -\partial_{\nu} A_{\mu}.
\end{eqnarray}
In the derivation of Eq.(\ref{action}),
one needs the classical equation of motion for the matter field.

The string variable $X^{\mu}$, having free (Neumann) boundary condition for
$\mu=m=0, \cdots ,p$, and fixed (Dirichlet) boundary condition for
$\mu=i=p+1, \cdots ,9$, satisfies
\begin{equation}
\partial_{\sigma}X^{m}\vert_{\sigma= 0,~\pi} = 0,~~\mbox{and}~~
\partial_{\tau}X^{i}\vert_{\sigma= 0, ~\pi} = 0, \end{equation} from which
we read the well-known fact in the oscillation problem of chords, i.e.,
the phase of oscillation does not change at the free boundary, whereas it
changes by $\pi$ at the fixed boundary. This fact is expressed as
$\alpha^{\mu}_{n}=\tilde{\alpha}^{\mu}_{n}$ for the Neumann boundary
condition, whereas $\alpha^{\mu}_{n}=-\tilde{\alpha}^{\mu}_{n}$ for the
Dirichlet boundary condition, since $\alpha^{\mu}_{n}$ and
$\tilde{\alpha}^{\mu}_{n}$ are right and 
left movers, which
correspond to the incoming (outgoing) and the outgoing (incoming)
oscillation modes, respectively.

More explicitly we have the following mode expansion: \begin{eqnarray}
X^{m}(\tau, \sigma)&=&x^{m} + \frac{\kappa}{\pi} p^{m} \tau + i
\sqrt{\frac{\kappa}{\pi}} \sum_{n \neq 0} \frac{\alpha^{m}_{n}}{n}
e^{-in\tau}2 \cos n\sigma, \\
X^{i}(\tau, \sigma)&=&x^{i} +\frac{\kappa}{\pi} \omega^{i} \sigma + i
\sqrt{\frac{\kappa}{\pi}} \sum_{n \neq 0} \frac{\alpha^{i}_{n}}{n}
e^{-in\tau} 2i\sin n\sigma,
\end{eqnarray}
where $p^{m}$ is center-of-mass coordinate, and $\omega^{i}$ is the winding
vector.  We take $\omega^{i}=0$ when both ends of the open string attach in
the neighborhood of the same D-brane, otherwise we need the
nonvanishing winding vector. 

Since the end points of the open string can be located on the D-brane,
various worldsheets of open strings can appear as the quantum
fluctuations.  Then, the external fields such as graviton, antisymmetric
tensor field, etc. may couple to these fluctuated worldsheets of the open
string, so that these external fields can be considered to interact with
the original D-brane indirectly.  If we integrate out the string variables,
the messenger of the interaction between D-brane and the external fields,
we will obtain the effective action of D-brane in the external fields. 
Therefore, open string worldsheets located apart from the world-volume
of the D-brane, do not contribute to the effective action of the D-brane,
even though they interact with the external fields. It is then sufficient to
take into account, as the messenger of the interaction, the string
coordinates inside the narrow strip sticking directly to the D-brane, and
to integrate them out.  Now, we can understand that apart from a constant
factor, the effective action so obtained does not depend on the shape of
the worldsheet, since the essential configuration is the narrow strip
sticking to the D-brane, but is irrelevant to the topology of the worldsheet 
of string as a whole.  Probably the coincidence of the effective
Born-Infeld actions up to a constant factor between tree and one-loop labels
found in Ref.\cite{F-T} comes from this property.

The narrow strip may be specified by 
$-T/2 <\tau< +T/2 ~( T\sim \infty)$,
$0<\sigma<\varepsilon \ll 1$, 
the string coordinates in which we are going 
to integrate out. 
In performing the path integration over 
the string coordinates on the narrow strip, we will
restrict the integration variables to the particular modes, satisfying
equation of motions like Eq.(5)-(7).  
This restriction corresponds
to the primary order of WKB approximation in the expansion of $\kappa$, or
the expansion in terms of the momentum $p^{m}$.  
Therefore, the
approximation is valid in deriving low energy effective action for
D-branes.
Then, we have 
\begin{eqnarray}
\label{epsaction}
\int^{+\infty}_{-\infty} d\tau \int^{\varepsilon}_{0} ( \eta^{ab} +
\epsilon^{ab} )~ \partial_{a} X^{\mu} \partial_{b} X^{\nu}&=&
\int^{+\infty}_{-\infty} d\tau \int^{\varepsilon}_{0} (\partial_{\tau}
-\partial_{\sigma}) X^{\mu} (\partial_{\tau} +\partial_{\sigma}) X^{\nu}\\
&=&T \varepsilon \frac{\kappa}{\pi} \sum^{\infty}_{-\infty} \pm
\alpha^{\mu}_{n} \alpha^{\nu}_{-n}, \label{oscillation modes}
\end{eqnarray}
where the signature $+$ and $-$ are associated with Neumann and Dirichlet
boundary conditions for the direction $\nu$, respectively.

In Eq.(\ref{oscillation modes}), we can see that the zero modes
$\alpha^{\mu}_{0}$ do not contribute. The zero modes are of course the
center of mass momentum and the winding (or length) vector,
\begin{equation}
\alpha^{\mu}_{0}= \left\{
\begin{array}{rl}
\frac{1}{2} \sqrt{\frac{\kappa}{\pi}} p^{m}~~&\mbox{for}~~\mu=m=0, \cdots
,p, \\ -\frac{1}{2} \sqrt{\frac{\kappa}{\pi}} w^{i}~~&\mbox{for}~~\mu=i=p;1,
\cdots ,9. \end{array}\right. \end{equation}
For the massless excitation of the open string, we may choose $w^{i}=0$,
namely the length of the string is vanishing. The meaning of the field
$N^{\mu}(X)$ can be seen from the variation of the ordinary string action
with respect to $X^{\mu}$,
\begin{equation}
\delta S = \frac{1}{\kappa} \int d \tau~~\delta X_{\mu}
\partial_{\sigma}X^{\mu}(\tau, \sigma=0). \end{equation} Therefore, the
variation of the end point $\delta X_{\mu}$, or the variation of the
D-brane's position is identified with the field $N_{\mu}(X)$, namely,
\begin{equation}
\label{Nposition}
N^{m}(x^0, \cdots, x^{p})=0, ~\mbox{and}~ N^{i}(x^0, \cdots,
x^{p})=Y^{i}(x^0, \cdots, x^{p}),
\end{equation}
where the first equation means that the variation in the tangential direction is
non-physical and its mode can be absorbed into the reparametrization of the
D-brane. Now, nonvanishing components of $N^{\mu\nu}$ are for $(\mu,
\nu)=( m, i)~\mbox{or}~(i, m)$. Then, we understand that the contribution
brane of the zero modes in Eq.(\ref{oscillation modes}) vanishes due to the
massless condition,
\begin{equation}
p^{m}G_{mn}(x)p^{n}=0.
\end{equation}
Now, we can perform the path integral over the open string modes within the
narrow strip,
\begin{eqnarray}
S_{eff}&=&\prod_{\mu=0}^{9} \int dx^{\mu} e^{-\phi(x)} \prod^{\infty}_{n=1}
\int d\alpha^{\mu}_{n} d\alpha^{\mu}_{-n}
e^{-T\varepsilon\frac{\kappa}{\pi} \sum^{\infty}_{n=1} \pm \left(
\alpha^{\mu}_{n} M_{\mu\nu}(x) \alpha^{\nu}_{-n}+ \alpha^{\mu}_{-n}
M_{\mu\nu}(x) \alpha^{\nu}_{n} \right) }\\ &=& \prod_{\mu=0}^{9} \int d
x^{\mu} e^{-\phi(x)} \left( \det_{\mu, \nu=0, \cdots ,9} M_{\mu\nu}(x)
\right)^{P}, \end{eqnarray}
where the power $P$ can be estimated by~\cite{F-T} \begin{equation}
P=\sum_{n=1}^{\infty} 1 = \left.\left( \sum_{n=1}^{\infty} n^{-s} \right)
\right|_{s \rightarrow 0} = \zeta(0) = - \frac{1}{2}. \end{equation} 

Then we obtain
\begin{equation}
S_{eff}= \int d^{p+1}x ~e^{-\phi(x)} \sqrt{ \det_{\mu, \nu=0, \cdots ,9}
\left( G_{\mu\nu}(x)+N_{\mu\nu}(x)+B_{\mu\nu}(x)+F_{\mu\nu}(x) \right) }.
\end{equation}
Here, the center-of-mass coordinates 
are restricted to those inside the D-brane.

In the effective action of the D-brane, the external fields coupling to it
must have their indices along its tangential direction as much as
possible. Here, we choose the flat background metric
$G_{\mu\nu}=\eta_{\mu\nu}$, the nonvanishing components of the two-form
field and the two vector fields as $B_{mn}(x^0, \cdots ,x^{p})$,
$A_{m}(x^0, \cdots ,x^{p})$, and $N_{i}(x^0, \cdots ,x^{p})$ given in
Eq.(\ref{Nposition}).

Then, we have\footnote{ This part follows the lecture given by N.
Ishibashi~\cite{Ishibashi} with a modification.}\ \begin{eqnarray}
\det_{\mu,\nu=0, \cdots ,9} M_{\mu\nu} &=& \det_{\mu, \nu} \left(
\begin{array}{c|c}
\eta_{mn}+B_{mn}+F_{mn} & \partial_{m}Y^{j}\\ \hline \partial_{n}Y^{i} &
\delta_{ij}
\end{array}
\right) \\
&=&
\det_{m, n=0, \cdots ,p} \left( \hat{G}_{mn} + {\cal F}_{mn} \right),
\end{eqnarray}
where $\hat{G}_{mn}$ is the induced metric on the D-brane, 
\begin{eqnarray}
\hat{G}_{mn}&=&\eta_{mn} - \sum^{9}_{i=p+1} \partial_{m}Y^{i}
\partial_{n}Y^{i} = \partial_{m}Y^{\mu} \eta_{\mu\nu} \partial_{n}Y^{\nu},
\\
{\cal F}_{mn}&=&B_{mn} - F_{mn}.
\end{eqnarray}

We have finally obtained the bosonic part $S_{eff}^{(1)}$ 
of the effective Born-Infeld
action for the D-brane, 
which is valid in the neighborhood of the D-brane,
\begin{equation}
S_{eff}^{(1)} = \int d^{p+1} x ~e^{-\phi(x)} \sqrt{ \det_{m, n=0, \cdots
,p} \left( \hat{G}_{mn}(x) + {\cal F}_{mn}(x) \right) }. 
\end{equation} 
%

\section{Born-Infeld Action for Superstrings} 

In order to generalize the rule of thumb derivation of Born-Infeld action
to superstrings, we have to add two more interactions: (a) interaction
of the fermionic string coordinate, $\psi^{\mu}(\tau, \sigma)$ with the
background fields of NS-NS sectors which appeared in the previous section.
(b) In addition, we have to introduce the interaction of the fermionic
string coordinate, $\psi^{\mu}(\tau, \sigma)$, with the Ramond-Ramond (R-R)
background fields. Those are the totally antisymmetric tensor fields with
$n$ indices ($n$-form fields), $A_{\mu_{1}, \mu_{2}, \cdots,
\mu_{n}}^{(n)}$, and $n$ is known to be odd and even for the type IIA and
type IIB superstring, respectively.

As for the interaction (a), using the worldsheet supersymmetry, the action
should be invariant under the following replacement, 
\begin{equation}
(\partial_{\tau}-\partial_{\sigma})X^{\mu}(\tau, \sigma) \rightarrow
\psi^{\mu}(\tau-\sigma), ~~\mbox{and}~~
(\partial_{\tau}+\partial_{\sigma})X^{\mu}(\tau, \sigma) \rightarrow
\tilde{\psi}^{\mu}(\tau+\sigma). \label{supersymmetry} 
\end{equation}
Therefore, we have the following effective action from Eqs. (\ref{action}) 
\begin{equation} S_{\psi}^{(1)}= \frac{1}{\kappa}\int
d\tau d \sigma~ M_{\mu\nu} \left( X(\tau, \sigma) \right)
\psi^{\mu}(\tau-\sigma) \tilde{\psi}^{\nu}(\tau+\sigma), 
\label{NS-NSinteraction} 
\end{equation} 
representing the interaction of the fermionic coordinates with the background 
NS-NS fields.

According to the supersymmetry in Eq.(\ref{supersymmetry}), the Neumann
and Dirichlet boundary conditions are transferred to the fermionic
coordinates, and we have $\psi^{\mu}_{k}=\pm \tilde{\psi}^{\mu}_{k}$, for
the $k$th oscillation modes. ($k$ is integer and half-integer for R and NS
sectors, respectively.)
Here, $+ (-)$ is chosen for the Neumann (Dirichlet) boundary condition,
giving the well-known fact of the phase shift 0 ($\pi$) at the free (fixed)
boundary of the chord as was discussed in the previous section. This naive
discussion is valid for the R sector, where the bosonic and fermionic
fields have both integer modes and the worldsheet supersymmetry is
manifest. In the NS-sector, however, the fermionic coordinate has
half-integer modes, different from the integer modes of the bosonic one, 
so that we cannot apply our naive argument, but then the local
supersymmetric generator $G_{\pm 1/2}$ with mode number $\pm 1/2$ plays
the role of the supersymmetry.

Here, in both NS and R sectors, we will adopt the following mode expansion
of the fermionic coordinates for $m=0, \cdots ,p$ and $i=p+1, \cdots ,9$:
\begin{eqnarray}
\left( \psi^{m}(\tau-\sigma), \tilde{\psi}^{m}(\tau+\sigma)
\right)&=&\sqrt{\frac{\kappa}{2\pi}} \sum_{k} \psi^{m}_{k} \left(
e^{-ik(\tau-\sigma)}, e^{-ik(\tau+\sigma)} \right), \\ \left(
\psi^{i}(\tau-\sigma), \tilde{\psi}^{i}(\tau+\sigma)
\right)&=&\sqrt{\frac{\kappa}{2\pi}} \sum_{k} \psi^{i}_{k} \left(
e^{-ik(\tau-\sigma)}, -e^{-ik(\tau+\sigma)} \right). \end{eqnarray} 

Next we will discuss the second interaction (\ref{action}) with the R-R
fields. As was found
by Polchinski~\cite{Polchinski}, these fields may couple to the D-brane's
volume element,
\begin{equation}
\int_{\mbox{\tiny D-brane}} dy^{m_{1}} \cdots dy^{m_{p+1}}~ A_{m_{1} \cdots
m_{p+1}} (y_{1}, \cdots, y_{p}).
\end{equation}
Since there exist smaller D-branes within the D-brane~\cite{Douglas}, the
interaction (\ref{action}) becomes the sum of their contributions,
\begin{equation} S^{(2)}_{\psi} = \sum_{ n \leq p+1 } \int_{n\mbox{\tiny
-brane} ~\subseteq ~\mbox{\tiny D-brane}} dy^{m_{1}} \cdots dy^{m_{n}}~
A_{m_{1}, \cdots, m_{n}} (y_{1}, \cdots, y_{p}),
\end{equation}
where the D-brane's position is specified by the proper coordinate system
of \hfill \\
$y = \left( y^{1}, \cdots, y^{p}, y^{i}=0~ (i=p+1, \cdots , 9) \right)$. We
can redundantly multiply the extra volume element (infinite but constant)
$dy^{p+1}\cdots dy^{9}$ to the above equation and obtain \begin{equation}
S'^{(2)}_{\psi} = \sum_{n} \int dy^{\mu_{1}} \cdots dy^{\mu_{n}} dy^{p+1}
\cdots dy^{9} A_{\mu_{1}, \cdots, \mu_{n}} (y_{1}, \cdots, y_{p}).
\label{R-R interaction}
\end{equation}
Then, the position of the D-brane is automatically specified and $n$-branes
within the D-brane are naturally chosen, without the restriction of
$n$-brane $\subseteq$ D-brane.

The problem is: what is the $n$-form $dy^{\mu_{1}} \wedge \cdots dy^{\mu_{n}}$
on the D-brane ? In order to discuss the interaction of the D-brane with
the R-R fields, the $n$-form should be rewritten in terms of the fermionic
coordinates $\psi^{\mu} (\tau, \sigma)$. The one-form on the D-brane $dy^{m}~
(m=0, \cdots ,p)$ can be represented as the motion of the end point of the
open string from $y^{m}$ to $y^{m}+dy^{m}$, which is generated by the zero
mode momentum $p^{m}$ of the string. Similarly, the form $dy^{i}~(i=p;1,
\cdots ,9)$ pointing outwardly from the D-brane is generated by the winding
(or the length) vector $w^{i}$ of the string. Since the zero mode momentum
and the winding vector is written as
$\alpha^{m}_{0}+\tilde{\alpha}^{m}_{0}$ and
$\alpha^{i}_{0}-\tilde{\alpha}^{i}_{0}$, respectively, they are transferred
to fermionic coordinates by the worldsheet supersymmetry: \begin{eqnarray}
dy^{m} \Leftrightarrow p^{m} &\rightarrow & \theta^{m}_{0} \equiv
\psi^{m}_{0}+\tilde{\psi}^{m}_{0}, \\
dy^{i} \Leftrightarrow w^{i} &\rightarrow& \theta^{i}_{0} \equiv
\psi^{i}_{0}-\tilde{\psi}^{i}_{0}.
\end{eqnarray}

Therefore, we can use $\theta^{\mu}_{0}$ as the one-form on the D-brane, and
$\theta^{\mu_{1}}_{0} \cdots \theta^{\mu_{n}}_{0}$ as the $n$-form on the
D-brane.

Now the interaction (\ref{action}) of the R-R fields with the D-brane,
given in Eq.(\ref{R-R interaction}), is rewritten as 

\begin{equation}
S''^{(2)}_{\psi}= \int d\tau d\sigma \sum_{n}~ \theta^{ \mu_{1}}_{0} \cdots
\theta^{\mu_{n}}_{0}~ \theta^{p+1}_{0} \cdots \theta^{9}_{0} ~\frac{1}{n!}
A^{(n)}_{\mu_{1}, \cdots, \mu_{p}} \left( X(\tau, \sigma) \right).
\end{equation}

Now, from Eqs.(\ref{NS-NSinteraction}) and (\ref{R-R interaction}), we
have the fermionic string action coupled to the external fields of both
types NS-NS and R-R as
\begin{equation}
S_{\psi}= S^{(1)}_{\psi} + S''^{(2)}_{\psi}. \end{equation} 

If we integrate out the fermionic coordinates $\psi^{\mu}(\tau, \sigma)$ in
the narrow strip \hfill\\
$(-T/2<\tau<T/2, 0< \tau \ll \varepsilon)$, we will obtain the effective
action for the D-brane.

First, using the identity,
\begin{equation}
\int \prod_{\mu=0}^{9} d\theta^{\mu}_{0}~ \theta^{\mu_{0}}_{0} \cdots
\theta^{\mu_{9}}_{0} = \epsilon^{\mu_{0}, \cdots, \mu_{9}}, \end{equation}
and noting the zero mode of $\psi^{\mu}(\tau-\sigma)
\tilde{\psi}^{\nu}(\tau+\sigma)$ to be $\frac{1}{2} \theta^{\mu}_{0}
\theta^{\nu}_{0}$, integration over zero modes $\theta^{\mu}_{0}$ gives the
following effective action,
\begin{eqnarray}
&S^{(2)}_{eff}&=
\int \prod_{\mu=0}^{9} d\theta^{\mu}_{0}~ e^{-S_{\psi}} \hfill \nonumber \\
&=& \int d^{p+1}x~ \epsilon^{\mu_{0} \cdots \mu_{9}} \Bigl(
\frac{1}{(p+1)!}~ A^{(p+1)}_{\mu_{0}, \cdots, \mu_{p}} (x) -
\frac{1}{(p-1)!}~ \frac{1}{2}~A^{(p-1)}_{\mu_{0}, \cdots, \mu_{p-2}}
(x)~{\cal F}^{(2)}_{\mu_{p-1}, \mu_{p}} (x) \nonumber \\
&+&\frac{1}{(p-3)!}~\frac{1}{8}~A^{(p-3)}_{\mu_{0}, \cdots, \mu_{p-4}} (x)~
{\cal F}^{(2)}_{\mu_{p-3}, \mu_{p-2}} (x) ~{\cal F}^{(2)}_{\mu_{p-1},
\mu_{p}} (x) + \cdots
\Bigr) \\
&=&\int d^{p+1}x ~ \left( \sum_{n~\leq~p+1} A^{(n\mbox{\tiny -form})}~
e^{-\frac{1}{2} {\cal F}^{(\mbox{\tiny two-form})} }
\right)_{((p+1)\mbox{\tiny -form})}. \label{WZ-term} \end{eqnarray}

Next, we will perform the nonzero mode integration over $\alpha^{\mu}_{\pm
n}$ and $\psi^{\mu}_{\pm k}$. As was found in Ref.\cite{Callan}, the bosonic
and fermionic mode integrations completely cancel in the R-sector, giving
no contribution to the effective action. Whereas in the NS-sector, bosonic
mode integration gives the effective action $S^{(1)}_{eff}$ given in
Eq.(22), but the fermionic mode integration gives nothing since
\begin{eqnarray}
\sum_{k=\mbox{\tiny positive half-integer}}~1~ &=&
2^{s}~\sum_{n=1}^{\infty}~\left( n^{-s} - (2n)^{-s} \right)
\vert_{s~\rightarrow~0} \\ &=& (2^{s}-1) \zeta(s) \vert_{s~\rightarrow~0} =
0. \end{eqnarray}
Usefulness of the zeta function regularization in the dynamics of extended
objects may be found in Ref.\cite{Odintsov}.

From the above discussions, we have finally obtain the Born-Infeld like
effective action for D-brane in the superstrings: 
\begin{eqnarray}
S_{eff} &=&S_{eff}^{(1)}+ S_{eff}^{(2)} \hfill \\ &=& \int d^{p+1} x \Bigl(
e^{-\phi(x)} \sqrt{ \det_{m, n=0, \cdots ,p} \left( \hat{G}_{mn}(x) +
{\cal F}_{mn}(x) \right) } \\
&+& ( \sum_{n~\leq~p+1} A^{(n)}~ \left.
e^{-\frac{1}{2} {\cal F}^{(2)}} )
\right|_{(p+1)\mbox{\tiny -form}} \Bigr).
\end{eqnarray}



\section{Conclusion}
Here we have discussed the rule of thumb derivation of the
Born-Infeld action for D-branes in the supersymmetric models. The
``derivation" is simply performed by integrating out of the superstring
coordinates in the narrow strip attached to the D-brane, so that it can be
viewed as a way to understand how the Born-Infeld action is derived. The
way of thinking is, however, easy to follow. Thus, this may
be helpful for us to study various problems in D-branes.

\smallskip

{\large\bf Note added} 
\\
In the paper by A.Tseytlin \cite{TT}, 
a similar idea of deriving explicitly the Born-Infeld 
action for D-brane has been given. 
In the superstring case, however, our treatment may 
be more direct than his. We are most grateful to 
Prof. A.A.Tseytlin
for pointing out his interesting paper.

\section*{Acknowledgements}
One of us (A.S.) is grateful to the members of the High Energy Theory
Groups of Niigata and Yamagata Universities, in the joint workshop (Oct
31-Nov. 2) of which, the content of this letter was presented. We would like to 
thank to Ryusuke Endo, Y. Igarashi, M. Imachi, T. Kuruma, T. Morozumi,
H. Nakano, T. Oonogi, T. Ozeki, and H. So for fruitful discussions.
We would also like to 
thank A.A.Tseytlin for informing us his publication and
Y.Kitazawa for reading the manuscript.



\begin{thebibliography}{8}
\bibitem{Polchinski} J. Polchinski, {\it Phys. Rev. Lett.} {\bf 75} (1995) 4724. 

\bibitem{F-T} E. S. Fradkin and A. A. Tseytlin, {\it Phys. Lett.} {\bf B163}
(1985) 123; {\bf B158} (1985) 316; {\it Nucl. Phys.} {\bf
B261} (1985) 1.



\bibitem{Cai} J. Polchinski and Y. Cai, {\it Nucl. Phys.} {\bf B296} (1988) 91. 

\bibitem{Callan} C. G. Callan, C. Lovelace, C. R. Nappi, and S. A. Yost,
{\it Nucl. Phys.} {\bf B308} (1988) 221.

\bibitem{Leigh}
J. Dai, R. G. Leigh, 
and J. Polchinski, {\it Mod. Phys. Lett.} {\bf A4} (1989) 2073;\\
R. G. Leigh, {\it ibid.} {\bf A4} (1989) 2767. 

\bibitem{Douglas} M. R. Douglas, hep-th/9512077. 

\bibitem{Li} M. Li, {\it Nucl. Phys.} {\bf B460} (1996) 351. 

\bibitem{Scmid} C. Schmidhuber, {\it Nucl. Phys.} {\bf B467} (1996) 146. 

\bibitem{Ishibashi} N. Ishibashi, Lecture given at Chubu Summer School at
Shiga Highland, Japan, August (1996).

\bibitem{Odintsov} See for example, E. Elizalde, S. D. Odintsov, A. Romeo,
A. A. Bytsenko, and S. Zerbini, {\it Zeta Regularization Techniques with
Applications} World Scientific (1994).

\bibitem{TT} A.A.Tseytlin, {\it Nucl.Phys.} {\bf B469}(1996)51.

\end{thebibliography}
\end{document}